\documentclass[11pt,prd,superscriptaddress,nofootinbib]{revtex4}
\usepackage[english]{babel}
\usepackage{graphicx}
\usepackage{mathtext}
\usepackage{indentfirst}
\usepackage{epsfig,amsmath,amsfonts,amssymb,braket}

\newcommand{\beq}[1]{\begin{eqnarray}\label{#1}}
\newcommand\eeq {\end{eqnarray}}
\newcommand\bqa {\begin{eqnarray}}
\newcommand\eqa {\end{eqnarray}}

\newcommand{\eq}[1]{(\ref{#1})}
\newcommand{\bear}{\begin{array}}
\newcommand{\enar}{\end{array}}

\begin{document}

\hfill ITEP--TH--39/14

\vspace{5mm}

\centerline{\Large \bf A few more comments on secularly growing loop corrections}
\centerline{\Large \bf in strong electric fields}

\vspace{5mm}

\centerline{E. T. ${\rm Akhmedov}^{1),3)}$ and F.K.${\rm Popov}^{2),3)}$}


\begin{center}
{\it $\phantom{1}^{1)}$ International Laboratory of Representation Theory and Mathematical Physics, National Research University Higher School of Economics, Russian Federation}
\end{center}

\begin{center}
{\it $\phantom{1}^{2)}$ Institutskii per, 9, Moscow Institute of Physics and Technology, 141700, Dolgoprudny, Russia}
\end{center}

\begin{center}
{\it $\phantom{1}^{3)}$ B. Cheremushkinskaya, 25, Institute for Theoretical and Experimental Physics, 117218, Moscow, Russia}
\end{center}

\vspace{3mm}

\begin{center}{\bf Abstract}\end{center}
We extend the observations of our previous paper JHEP 1409, 071 (2014)
[arXiv:1405.5285]. In particular, we show that the secular growth of the loop corrections to the two--point correlation functions is gauge independent: we observe the same growth in the case of the static gauge for the constant background electric field. Furthermore we solve the kinetic equation describing photon production from the background fields, which was derived in our previous paper and allows one to sum up leading secularly growing corrections from all loops. Finally, we show that in the constant electric field background the one--loop correction to the current of the produced pairs is not zero: it also grows with time and violates time translational and reversal invariance of QED on the constant electric field background.

\vspace{5mm}

\section{Introduction}

Schwinger's pair creation \cite{Schw} is a well studied phenomenon. However, in our recent paper \cite{Akhmedov:2014hfa} we show that in QED on strong electric field backgrounds there are loop corrections to propagators which grow with time. We use Schwinger--Keldysh diagrammatic technique and consider a constant electric field, $E_z = const$, and electric pulse, $E_z(t) \propto \frac{1}{\cosh^2{\left(t/T\right)}}$. We show that after a long enough evolution in a constant field background (or as $T\to \infty$ for the case of the pulse), loop corrections become of the order of the tree--level contribution, which substantially changes the picture of the particle production. That happens due to the secular growth of the loop corrections to propagators. This effect cannot be seen in the standard approaches to the subject (see e.g. \cite{Gitman1} -- \cite{Grib}), which are mostly applicable in the background field approximation (we come back to this point below in this section).

Before discussing the loopholes in the standard approaches let us explain the physical origin of the secular loop effects. The point is that secular growth of loop corrections is quite a generic situation as is know in condensed matter theory \cite{LL}, \cite{Kamenev}. To see that we start with the explanation of the reason why one has to apply the Schwinger--Keldysh technique instead of the Feynman one in non--stationary situations.
Suppose one would like to find the time evolution of the expectation value of an operator ${\cal O}$:

\bqa\label{ave}
\left\langle {\cal O} \right\rangle(t) \equiv \left\langle \Psi\left| \overline{T} e^{i\,\int_{t_0}^t dt' H(t')}\,{\cal O} \, T e^{-i\,\int_{t_0}^t dt' H(t')}\right| \Psi\right\rangle.
\eqa
Here $H(t) = H_0(t) + H^{int}(t)$ is the full Hamiltonian of a theory, $T$ denotes the time--ordering and $\overline{T}$ is the reverse time--ordering; $t_0$ is an initial moment of time and $\left|\Psi\right\rangle$ is an initial state. We assume that the initial value $\left\langle {\cal O} \right\rangle(t_0)$ is given.

After the transformation to the interaction picture, we get \cite{LL}:

\bqa\label{step}
\left\langle {\cal O} \right\rangle(t) = \left\langle \Psi\left| S^+(t, t_0)\, {\cal O}_0(t) \, S(t,t_0) \right| \Psi\right\rangle = \left\langle \Psi\left| S^+(t, t_0)\, T\left[ {\cal O}_0(t) \, S(t,t_0)\right]\right| \Psi\right\rangle = \nonumber \\ = \left\langle \Psi\left| S^+(t, t_0) S^+(+\infty, t)\, S(+\infty, t)\, T\left[ {\cal O}_0(t) \, S(t,t_0)\right]\right| \Psi\right\rangle = \nonumber \\ = \left\langle \Psi\left| S^+(+\infty, t_0)\, T\left[ {\cal O}_0(t) \, S(+\infty,t_0)\right]\right| \Psi\right\rangle,
\eqa
where $S(t,t_0) = T e^{-i\,\int_{t_0}^t dt' H^{int}_0(t')}$; ${\cal O}_0(t)$ and $H^{int}_0(t)$ are the same operators as above, but written in the interaction picture. To perform the first step in (\ref{step}) we have used the Baker-Hausdorff formula. To perform the step on the second line of (\ref{step}) we had inserted the following resolution of the unit operator: $1 = S^+(+\infty, t)\, S(+\infty, t)$. That allows one to extend the original evolution (from $t_0$ to $t$, $S$ and back, $S^+$) to that which goes from $t_0$ to future infinity, $S(+\infty, t_0)$, and back, $S^+(+\infty, t_0)$. We put the operator ${\cal O}_0(t)$ on the forward going part, $S$, of the time contour.

To convert (\ref{step}) into a suitable form we assume that interactions, $H^{int}$, are adiabatically turned on after $t_0$, i.e., $\left|\Psi \right\rangle$ does not evolve before $t_0$. Then, one can rewrite the expectation value (\ref{step}) as follows:

\bqa\label{calo}
\left\langle {\cal O} \right\rangle_{t_0}(t) = \left\langle \Psi\left| S_{t_0}^+(+\infty, -\infty)\, T\left[ {\cal O}_0(t) \, S_{t_0}(+\infty,-\infty)\right]\right| \Psi\right\rangle.
\eqa
A good question is if one can take $t_0$ to past infinity, $t_0 \to -\infty$, i.e. to get rid of the dependence of $\left\langle {\cal O} \right\rangle_{t_0}(t)$ on $t_0$. The seminal example when one can do so is as follows: The free Hamiltonian, $H_0$, does not depend on time and $\left|\Psi \right\rangle$ coincides with its ground state $\left|vac \right\rangle$, $H_0 \, \left|vac \right\rangle = 0$. One also assumes that the interaction term is adiabatically switched off at future infinity --- long after the time $t$.

If $\left|vac \right\rangle$ is the true vacuum state of the free theory, then, by adiabatic turning on and then switching off the interactions, one cannot disturb such a state, i.e., $\left\langle vac \left| S^+(+\infty, -\infty)\right| excited \,\, state\right\rangle = 0$, while $\left|\left\langle vac \left| S^+(+\infty, -\infty)\right| vac \right\rangle\right| = 1$. Then the dependence on $t_0$ disappears. Hence,

\bqa\label{43}
\left\langle {\cal O} \right\rangle(t) = \sum_{state} \left\langle vac \left| S^+(+\infty, -\infty)\right| state \right\rangle \, \left\langle state \left| T\left[ {\cal O}_0(t) \, S(+\infty,-\infty)\right]\right| vac \right\rangle = \nonumber \\ = \left\langle vac \left| S^+(+\infty, -\infty)\right| vac \right\rangle \, \left\langle vac \left| T\left[ {\cal O}_0(t) \, S(+\infty,-\infty)\right]\right| vac \right\rangle = \nonumber \\ =
\frac{\left\langle vac \left| T\left[ {\cal O}_0(t) \, S(+\infty,-\infty)\right]\right| vac \right\rangle}{\left\langle vac \left| S(+\infty, -\infty)\right| vac \right\rangle} .
\eqa
To perform the first step in (\ref{43}), we have inserted the resolution of unity $1 = \sum_{state} \left| state \right\rangle \, \left\langle state \right|$, where the sum is going over the complete basis of eigen-states of $H_0$. To perform the second step, we have used that $\left|vac \right\rangle$ is the only state from the sum which gives a non-zero contribution. Thus, we arrive at the expressions which contain only T-ordering (and no any $\overline{\rm T}$-orderings), i.e., we obtain the standard Feynman diagrammatic technique. In this case one can shift the moment after which interactions are adiabatically turned on to past infinity, $t_0 \to - \infty$.

However, if $\left|\Psi\right\rangle$ is not a ground state and/or $H_0$ depends on time, one cannot use the above machinery and has to deal directly with \eq{calo} or (\ref{step}). In this case the efficient method is the so-called Schwinger--Keldysh technique, where one has to perturbatively expand both $S$ and $S^+$ under the quantum average.
Then each vertex in the expansion comes either from $S$ (and then assigned ``$+$'' sign) or --- from $S^+$ (and then assigned the ``$-$'' sign).

As follows from these considerations Schwinger--Keldysh technique is causal, unlike the Feynman one, and is used without any appeal to the notion of particle. In fact, with this technique one calculates correlation functions rather than S--matrix elements, i.e. one does not need to define what are asymptotic states. The latter ones are ambiguous if a background field is not switched off, because the free Hamiltonian is never diagonal. Finally note that one can also apply the Schwinger--Keldysh technique in the stationary situation because then the $\overline{\rm T}$-ordered expressions just cancel out vacuum diagrams. Many comparatively simple and interesting examples of the application of this technique are presented in \cite{Kamenev}.

Now we are ready to explain the origin of the secularly growing loop corrections in non--stationary situations. Basically in any propagator there is such an element as $\langle a^+ a \rangle$, where $a$ and $a^+$ are annihilation and creation operators for a field under consideration. In the standard Feynman technique the average is done with the use of the ground state, hence, $a |0\rangle = 0$ and this element is vanishing: If the background state is true vacuum then $\langle a^+ a \rangle$ is zero from the very beginning and remains zero at future infinity. If, however, the background state is not a true vacuum, then the situation is drastically different. However, even in a non--stationary situation at tree--level $\langle a^+ a \rangle$ remains constant, if all the time dependence is absorbed into exact harmonics. (This is the case in the interacting picture.) But if one turns on interactions then $\langle a^+ a \rangle$ starts to depend on time. It starts to run immediately right after $t_0$ --- the moment after which the interactions are adiabatically turned on. In a generic situation one finds that $\langle a^+ a \rangle \propto (t-t_0)$, where the coefficient of proportionality is the collision integral, which is not zero because the situation is not stationary. The linear growth appears, if the collision integral is constant in time.
That is the reason why in such a case one cannot take $t_0$ to the past infinity, if he keeps the population numbers fixed in the loop corrections --- in the calculation of the collision integral (see e.g. \cite{LL}).

Such a growth of the loop corrections to the two--point correlation functions has bright physical consequences.
In particular in \cite{Akhmedov:2014hfa}, we observe that particle number density, $\langle \alpha^+_{\mu p} \alpha^\mu_{p}\rangle$, which is an element of the photon's Keldysh propagator, grows with time even if at the initial state it was zero. Thus, there is photon production together with charged particles from the background electric field. (If interactions between quantum charged and gauge fields are turned on, photons are produced by the background field together with the charged particles rather than by accelerating products of the pair creation, i.e. photons are produced even if the density of the charged pairs is zero.) This is true for the both types of electric backgrounds under consideration --- constant and pulse.

Note that these observations explain the following controversy in the constant electric field background. It happens that in this case the decay rate is not zero, but the current of the created particles does vanish \cite{Anderson:2013zia}, \cite{Anderson:2013ila}, due to the symmetries of the problem. The dependence of the correlation functions on $t_0$ brakes the time--reversal invariance of the QED on the constant field background and allows to have a non--zero current in the loops, which explains this puzzle. We present the details of this explanation at the end of the present paper.

It is worth pointing out now that secular growth of loop corrections is observed also in the case of other strong background fields: See \cite{PolyakovKrotov}, \cite{Akhmedov:2008pu}, \cite{Akhmedov:2009be}, \cite{AkhmedovKEQ}, \cite{AkhmedovBurda}, \cite{Polyakov:2012uc}, \cite{AkhmedovGlobal}, \cite{AkhmedovPopovSlepukhin} for the same kind of effects in de Sitter space and \cite{AkhmedovReview} for a review.

Let us continue our discussion with the critique of the standard approaches and explaining the reason why one cannot see the secular growth of the correlation function, if one applies them. The original calculation goes as follows: one looks for the ground states of the free Hamiltonian in the background electric field at past and future infinity --- $|in\rangle$ and $|out\rangle$, correspondingly. In this case one finds that, unlike the case of empty space quantum field theory, $\left|\langle in | out\rangle\right| \neq 1$. Namely, it happens that $\langle in | out\rangle = e^{- \Gamma\, V}$, where $V$ is the four--volume. Here $\Gamma$ is interpreted as a probability rate per unit four--volume for the decay of the ground state. Furthermore, in the Feynman loop calculations within this context one finds peculiarities (such as imaginary contributions) of the effective actions in the strong electric field backgrounds. In all such calculations photon field is kept non--dynamical background field. Some of these effects one can grasp via the analytical continuation from the Euclidian space instantons.

But one can ask the following questions: Does the ground state indeed decay only via pair creation? Is it really true that all other kinds of processes do not appear during the whole period of evolution of the system under consideration from past to future infinity? Perhaps the ground state does indeed decay only via the pair creation if one considers the pulse background and takes the limit when electric change is tending to zero, $e\to 0$, background field is tending to infinity, $E \to \infty$, while their product is kept finite, $eE = const$. But then how one can calculate corrections to this picture beyond the latter limit? How one can go beyond the background field approximation?

In principle one can calculate rates of other types of more complicated creation processes with the use of Feynman technique. In fact, it seems that one just has to consider more complicated amplitudes rather than just the simplest one $\langle in | out\rangle$. But what about loop corrections to these amplitudes? The problem is that due to the use of exact harmonics instead of plane waves there is no energy conservation in the vertexes in the diagrams. Because of that IR loop corrections do not factor out and the standard cancelation of the IR divergences does not work \cite{Akhmedov:2009vh}. Apart from that the standard cancelation of IR divergencies goes via a redefinition of the asymptotic states. But how should one redefine the states, if at the past infinity harmonic functions behave as $e^{-i \,\omega(p) \, t}$, while at the future --- as the linear combinations $\alpha e^{- i \omega(p) t} + \beta e^{i \omega(p) t}$? As the result of these IR divergencies all cross sections are either zero or infinite and the calculation of the standard S--matrix elements is just meaningless.

This is not so surprising if one realizes that the system under consideration is not closed, as should be the case in strong background fields. Note that all the above observations are in effect only at the non--perturbative level in the background field and cannot be seen at any finite order in the expansion over the field: We use the exact harmonics and propagators.

Finally note that in \cite{Barashev:1985zm} they calculate photon's self--energy using time--ordered in--in propagators. To see the effects that we are after in this note one has to calculate correction to the photon's propagator rather than just the photon's self--energy. Moreover, the time--ordered contributions of \cite{Barashev:1985zm} bring only one $(++)$ correction, in the Schwinger's notations. While the complete correction to the propagator includes the sum of all $(++)$, $(--)$, $(+-)$ and $(-+)$ contributions.

The purpose of this note is as follows. In \cite{Akhmedov:2014hfa} we have shown that, unlike the case of the Keldysh propagator, the photon's retarded and advanced propagators and all propagator's of the charged particles do not receive such secularly growing corrections at the first loop. Also vertexes do not receive corrections that grow with time. These observations allowed us to simplify the system of the Dyson--Schwinger equations to take into account leading corrections. Along these lines we derive in \cite{Akhmedov:2014hfa} the kinetic equation for the photon production in the strong field backgrounds. The solution of this equation allows one to sum up leading secularly growing corrections from all loops.

All these observations in \cite{Akhmedov:2014hfa} have been made in the $A_\mu = \left(0,0,0, A_3 = Et\right)$ gauge in the case of the constant field background. One of the goals of the present paper is to show that our result is gauge independent, i.e. we would like to repeat the calculation in the $A_\mu = \left(A_0 = - Ez, 0,0,0\right)$ gauge. The point is that the observations of \cite{Akhmedov:2014hfa} are based on the fact that there is no energy conservation in time dependent backgrounds. Hence, it may seem unclear what the reason for the same phenomenon in the static gauge under consideration is, once there is energy conservation for the single particle problem. In this note we clarify this point.

Another goal of this note is to solve the aforementioned kinetic equation for photons. And finally we would like to see the impact of these effects on the current of the produced charged particles. The point is that at tree--level this current is zero in the constant electric field background \cite{Anderson:2013zia}, \cite{Anderson:2013ila}, because of the invariance of QED under the time translational and reversal invariance on the eternally and everywhere constant field background. We show that at the loop order these symmetries are broken and the current receives non--zero contributions that grow with time.

\section{Setup of the problem}

We consider, here, a massive scalar field coupled to an electromagnetic field in $(3+1)$ dimensions:

\begin{gather*}\label{action}
S = \int d^4 x\left[|D_\mu \phi|^2 - m^2 |\phi|^2 - \frac{1}{4} F_{\mu\nu}^2 - j^{cl}_\mu A^\mu\right],
\end{gather*}
where $D_\mu = \partial_\mu - i e A_\mu$. We divide the full gauge potential into two pieces $A_\mu = A^{cl}_\mu + a_\mu$ --- classical, $A_\mu^{cl}$, and quantum, $a_\mu$, parts. Throughout this paper, we denote the external gauge-potential $A^{cl}_\mu$ as $A_\mu$. If not otherwise stated, in this note we study the constant field background in static gauge, where $A_0(z) =  - E z$ and $\vec{A} = 0$.

The quantization of the gauge field is straightforward. One just has to choose a convenient gauge for $a_\mu$. Below we choose Feynman gauge. For the charged scalars the situation is not so transparent because we use exact harmonics in the background field rather than plane waves. So we give here a few comments on how to quantize the theory in such a situation.

Introducing the following notations $\bar{k} = \left(k_0,k_1,k_2\right)$, $\vec{k}_\perp = (k_1,k_2)$ and $d^3\bar{k} = dk_0 d^2 \vec{k}_\perp$, we expand the charged scalar fields in harmonics as follows:

\begin{eqnarray} \label{harmexp}
\phi(x,t) = \int \frac{d^3\bar{k}}{(2\pi)^3} \, \left\{ a_{\bar{k}} f_{k_\perp}\left(z - \frac{k_0}{e E}\right) \, e^{- i k_0 t + i \vec{k}_\perp \cdot \vec{x}_\perp} + b^\dagger_{\bar{k}}f^*_{k_\perp}\left(- z - \frac{k_0}{e E}\right) \,  e^{i k_0 t - i \vec{k}_\perp \cdot \vec{x}_\perp}\right\}.
\end{eqnarray}
The function $f_{k_\perp}\left(z - \frac{k_0}{e E}\right)$ satisfies the following differential equation:

\begin{eqnarray}
- \left[\left(\partial_t + i e E z\right)^2 - \vec{\partial}_\perp^2 - \partial_z^2 + m^2\right] f_{k_\perp}\left(z - \frac{k_0}{e E}\right) \, e^{- i k_0 t + i \vec{k}_\perp \cdot \vec{x}_\perp} = \nonumber \\ = \left[\partial_z^2 + \left(k_0 -  e E z\right)^2 - k_\perp^2 - m^2\right] f_{k_\perp}\left(z - \frac{k_0}{e E}\right) \, e^{- i k_0 t + i \vec{k}_\perp \cdot \vec{x}_\perp} = 0.\label{eqEz}
\end{eqnarray}
Solutions of (\ref{eqEz}) are related via a Fourier transformation, which we give below, to those of:

\begin{equation}
\left[\partial_t^2 + \left(k_3 +  e E t\right)^2 + k_\perp^2 + m^2\right] f_{k_\perp}\left(t + \frac{k_3}{eE}\right) = 0.\label{eqE}
\end{equation}
This equation defines harmonic functions in the temporal, $A_3 = Et$, gauge (see e.g. \cite{Akhmedov:2014hfa}). The Fourier relation in question can be seen after the change of variables $k_0 - eEz = - eEZ$ and $ eET = k_3 + eEt$. Then the solutions of (\ref{eqEz}) and (\ref{eqE}) are related as follows:

\begin{equation}\label{Fourier}
\int_{-\infty}^{+\infty} dT \, f_{k_\perp}\left(T\right) \, e^{- i e E T Z} = f_{k_\perp}\left(Z\right)
\end{equation}
We use this Fourier relation throughout the paper and we give the explicit form of $f_{k_\perp}$ below.

From the commutation relations $\left[a_{\bar{k}},a^\dagger_{\bar{k}'}\right] = \left[b_{\bar{k}},b^\dagger_{\bar{k}'}\right] = \left(2\pi\right)^3 \delta^{(3)}(\bar{k} - \bar{k}')$ the commutation relations between $\phi$ and its conjugate momentum $\pi = \left(\partial_t - i e E z\right)\phi^*$ takes the standard form:

\begin{eqnarray*}
\left[\phi\left(t,\vec{x}_1\right), \, \pi\left(t,\vec{x}_2\right)\right] =i  \int \frac{d^3\bar{k}}{(2\pi)^3} \, \left(k_0 - e E z_2\right) e^{ i \vec{k}_\perp \cdot \left(\vec{x}_{1\perp} - \vec{x}_{2\perp}\right)} \times \nonumber \\ \times \left[f_{k_\perp}\left(z_1 - \frac{k_0}{e E}\right) \, f^*_{k_\perp}\left(z_2 - \frac{k_0}{e E}\right) - f^*_{k_\perp}\left(- z_1 + \frac{k_0}{e E}\right) \, f_{k_\perp}\left(- z_2 + \frac{k_0}{e E}\right)\right] = i \delta^{(3)}\left(\vec{x}_1 - \vec{x}_2\right)
\end{eqnarray*}
The last equality follows from the Fourier transformation (\ref{Fourier}). Also one has to use the conservation of the Wronskian for the solutions of (\ref{eqE}).

The free Hamiltonian for the charged scalars is diagonal:

\begin{eqnarray}
H_0 = \int d^3 x \left[|\partial_t \phi|^2 + |\partial_i \phi|^2 + m^2 |\phi|^{2\phantom{\frac12}} - e^2 E^2 z^2 |\phi|^2\right] = \notag\\ = \int d^3 x \left[|\partial_t \phi|^2 - \phi^* \partial_t^2 \phi^{\phantom{\frac12}} - 2 i e E z \phi^* \partial_t \phi\right]
= \int \frac{d^3\bar{k}}{(2\pi)^3} \, k_0 \, \left[a^\dagger_{\bar{k}}a_{\bar{k}}+b^\dagger_{\bar{k}}b_{\bar{k}}\right].
\end{eqnarray}
Here we have used the harmonic expansion of $\phi$ and the fact that harmonic functions obey the equation of motion, (\ref{eqEz}).

From the obtained form of the free Hamiltonian (it is diagonal and time independent) we can see that in static gauge in the constant electric field background there is energy conservation for each single harmonic. But the energy is not bounded from below, because $k_0$ can have any sign. Because of the latter fact we will see that various particle creation processes will be allowed when the interaction with the quantum gauge field, $a_\mu$, will be turned on.

\section{One--loop correction}

Because the free Hamiltonian, $H_0$, is not bounded from below, the field theory under consideration is in the non-stationary situation. Hence, to calculate correlation functions one has to apply the Keldysh-Schwinger (KS) diagrammatic technique instead of the Feynman one \cite{LL}, \cite{Kamenev}. In such a formalism every particle is described by the matrix propagator, whose entries are the Keldysh propagator $G^K_{\mu\nu} = \frac{1}{2} \left\langle\left\{a_\mu(x), a_\nu(y)\right\}\right\rangle$, and the retarded and advanced propagators $G_{\mu\nu}^{A,R} = \mp\theta(\mp\Delta t)\left\langle\left[a_\mu(x), a_\nu(y)\right]\right\rangle$ (and the same for the scalar fields, with $a_\mu \to \phi$).

For our discussion it is instructive to see how the Keldysh propagators behave if the quantum average is done with the use of an arbitrary state $|\psi \rangle$. Performing the harmonic expansion of the quantum part, $a_\mu(x)$, of the photon field

$$a_\mu(x) = \int \frac{d^3\vec{q}}{(2\pi)^3 \, \sqrt{2|q|}} \left(\alpha_{\vec{q}\mu} \, e^{- i|q| t + i \vec{q} \cdot \vec{x}} + h.c.\right),$$
we find that the photon's Kledysh propagator has the following form:

\begin{eqnarray}\label{refpoint}
G^K_{\mu\nu}(x_1,x_2) =  \frac12 \, \int \frac{d^3\vec{q} d^3\vec{q}'}{(2\pi)^6} \, \left\{n_{\mu\nu}\left(\vec{q}, \vec{q}'\right) \, \frac{e^{i q \cdot x_1 - i q' \cdot x_2}}{\sqrt{\left|\vec{q}\right|\,\left|\vec{q}'\right|}} +  \kappa_{\mu\nu}\left(\vec{q}, \vec{q}'\right) \, \frac{e^{- i q \cdot x_1 - i q' \cdot x_2}}{\sqrt{\left|\vec{q}\right|\,\left|\vec{q}'\right|}} + h.c.\right\}.
\end{eqnarray}
Here  $n_{\mu\nu}(\vec{q}, \vec{q}') = \left\langle\psi\left|\alpha^\dagger_{\vec{q}\mu} \alpha_{\vec{q}'\nu}\right|\psi\right\rangle$, $\kappa_{\mu\nu}(\vec{q}, \vec{q}') = \left\langle\psi\left|\alpha_{\vec{q}\mu} \alpha_{\vec{q}'\nu}\right|\psi\right\rangle$ and $q\cdot x = |q| t - \vec{q} \cdot \vec{x}$. Furthermore, $h.c.$ stands for the quantities containing $\left\langle\psi\left|\alpha_{\vec{q}\mu} \alpha^\dagger_{\vec{q}'\nu}\right|\psi\right\rangle = n_{\mu\nu}(\vec{q}, \vec{q}') - g_{\mu\nu}\, \delta^{(3)}\left(\vec{q} - \vec{q}'\right)$ and $\kappa^*_{\mu\nu}\left(\vec{q}, \vec{q}'\right) = \left\langle\psi\left|\alpha^\dagger_{\vec{q}\mu} \alpha^\dagger_{\vec{q}'\nu}\right|\psi\right\rangle$.

Furthermore from (\ref{harmexp}) we find that scalar field's Keldysh propagator is as follows:

\begin{eqnarray}\label{loop2}
D^K(x_1,x_2) = \frac12 \left\langle \left\{\phi(x_1), \, \bar{\phi}(x_2)\right\}\right\rangle = \frac12 \, \int \frac{d^3\bar{k} d^3\bar{k}'}{(2\pi)^6} \times \nonumber \\ \times \left\{n^+\left(\bar{k}, \bar{k}'\right) \, e^{i k_0 t_1 - i \vec{k}_\perp \cdot \vec{x}_{1\perp}} \, e^{-i k'_0 t_2 + i \vec{k}'_\perp \cdot \vec{x}_{2\perp}} \, f^*_{k_\perp}\left(z_1 - \frac{k_0}{e E}\right) \, f_{k'_\perp}\left(z_2 - \frac{k'_0}{e E}\right)
+ \right. \nonumber \\ + \left. \kappa^+\left(\bar{k}, \bar{k}'\right) \, e^{- i k_0 t_1 + i \vec{k}_\perp \cdot \vec{x}_{1\perp}} \, e^{- i k'_0 t_2 + i \vec{k}'_\perp \cdot \vec{x}_{2\perp}} \, f_{k_\perp}\left(z_1 - \frac{k_0}{e E}\right) \, f_{k'_\perp}\left(- z_2 - \frac{k'_0}{e E}\right) + h.c. \right\}.
\end{eqnarray}
Here $n^+\left(\bar{k}, \bar{k}'\right) = \left\langle\psi\left|a^\dagger_{\bar{k}} a_{\bar{k}'}\right|\psi\right\rangle$, $\kappa^+\left(\bar{k}, \bar{k}'\right) = \left\langle\psi\left|a_{\bar{k}} b_{\bar{k}'}\right|\psi\right\rangle$
and $h.c.$ stands for the expressions containing $\left\langle\psi\left|a_{\bar{k}} a^\dagger_{\bar{k}'}\right|\psi\right\rangle = \delta^{(3)}\left(\bar{k} - \bar{k}'\right) + n^+\left(\bar{k}, \bar{k}'\right)$, $\left\langle\psi\left|b^\dagger_{\bar{k}} b_{\bar{k}'}\right|\psi\right\rangle = n^-\left(\bar{k}, \bar{k}'\right)$, $\left\langle\psi\left|b_{\bar{k}} b^\dagger_{\bar{k}'}\right|\psi\right\rangle = \delta\left(\bar{k} - \bar{k}'\right) + n^-\left(\bar{k},\bar{k}'\right)$ and $\kappa^-\left(\bar{k}, \bar{k}'\right) = \left\langle\psi\left|a^\dagger_{\bar{k}} b^\dagger_{\bar{k}'}\right|\psi\right\rangle$.

At the same time the form of the retarded and advanced propagators does not depend on the state
$|\psi \rangle$. In \cite{Akhmedov:2014hfa} it was shown that there are no large (growing with time) loop corrections to the retarded and advanced propagators and also to the vertexes. This is a quite generic phenomenon: see e.g. \cite{Kamenev} for the similar situations in different theories. It is straightforward to show that the same is true in the static gauge. Hence, we continue with the discussion of the Keldysh propagators. The reason why we present (\ref{refpoint}) and (\ref{loop2}) here is that loop corrections contribute to $n$ and $\kappa$ in the Keldysh propagators of both fields.

\subsection{Correction to the photon's Keldysh propagator}

We start with the one--loop correction to the photon's Keldysh propagator in the limit $\frac{t_1+t_2}{2} = t \to \infty$, when $t_1 - t_2 = {\rm const}$. The initial state that we consider here is the one that is annihilated by all annihilation operators under consideration ($a$'s, $b$'s and $\alpha$'s). I.e. the tree--level Keldysh propagators $G^K$ and $D^K$ look as (\ref{refpoint}) and (\ref{loop2}) with all $n$ and $\kappa$ equal to zero\footnote{Note that then $G^K$ and $D^K$ in (\ref{refpoint}) and (\ref{loop2}) are not zero because $\langle vac| \alpha_{\vec{q}\mu} \, \alpha^+_{\vec{q}'\nu} |vac\rangle = - g_{\mu\nu} \, \delta^{(3)}\left(\vec{q} - \vec{q}'\right)$ and $\left\langle ground \left| a_{\bar{k}} \, a^+_{\bar{k}'} \right| ground \right\rangle = \delta^{(3)}\left(\bar{k} - \bar{k}'\right)$.}.

Performing the same calculation as in \cite{Akhmedov:2014hfa} one can see that the one--loop correction to the propagator in question has the form of (\ref{refpoint}), where $n_{\mu\nu}\left(\vec{q}, \vec{q}', t\right) = \delta^{(2)}\left(\vec{q}_\perp - \vec{q}'_\perp\right) \, n_{\mu\nu}\left(q_3,q'_3, \vec{q}_\perp, t\right)$ and $\kappa_{\mu\nu}\left(\vec{q}, \vec{q}', t\right) = \delta^{(2)}\left(\vec{q}_\perp - \vec{q}'_\perp\right) \, \kappa_{\mu\nu}\left(q_3,q'_3, \vec{q}_\perp, t\right)$. The latter quantities are as follows:

\begin{eqnarray}\label{nkappa}
n_{\mu\nu}\left(q_3,q'_3, \vec{q}_\perp, t\right) \approx
e^2\int \frac{d^3\bar{k}}{(2\pi)^3} \, \int \frac{dk_0'}{2\pi}  \int\limits^{t}_{t_0}\int\limits^{t}_{t_0} dt_3 dt_4 \frac{e^{- i \left(k_0+k_0'\right) (t_3-t_4)} e^{- i |q| t_3 + i |q'| t_4}}{2 \sqrt{|q||q'|}}\times\notag\\\times
\int dz_3 e^{ i q_3 z_3} \left[f_{k_\perp}\left( z_3 - \frac{k_0}{e E}\right)\overleftrightarrow{D_\mu} f_{\left|\vec{q}_\perp + \vec{k}_\perp\right|}\left(-z_3 - \frac{k_0'}{e E}\right)\right]\times\notag\\\times
\int dz_4 e^{ - i q'_3 z_4} \left[f^*_{k_\perp}\left(z_4 - \frac{k_0}{e E}\right)\overleftrightarrow{D_\nu} f^*_{\left|\vec{q}_\perp + \vec{k}_\perp\right|}\left(-z_4 - \frac{k_0'}{e E}\right)\right], \notag\\
{\rm and} \quad \kappa_{\mu\nu}\left(q_3, q_3', \vec{q}_\perp,t\right) \approx - 2
e^2\int \frac{d^3\bar{k}}{(2\pi)^3}\, \int \frac{dk_0'}{2\pi}  \int\limits^{t}_{t_0}\int\limits^{t_3}_{t_0} dt_3 dt_4 \frac{e^{- i \left(k_0+k_0'\right) (t_3-t_4)} e^{- i |q| t_3 - i |q'| t_4}}{2 \sqrt{|q||q'|}}\times\notag\\\times
\int dz_3 e^{ i q_3 z_3} \left[f_{k_\perp}\left(z_3 - \frac{k_0}{e E}\right)\overleftrightarrow{D_\mu} f_{\left|\vec{q}_\perp + \vec{k}_\perp\right|}\left(-z_3 - \frac{k_0'}{e E}\right)\right] \times \notag \\ \times
\int dz_4 e^{i q'_3 z_4} \left[f^*_{k_\perp}\left(z_4 - \frac{k_0}{e E}\right) \overleftrightarrow{D_\nu} f^*_{\left|\vec{q}_\perp + \vec{k}_\perp\right|}\left(-z_4 - \frac{k_0'}{e E}\right)\right],
 \end{eqnarray}
where $D_\mu f_{p_\perp}\left(\pm z - p_0/eE\right) = \left(-ip_0 \pm i e E z, i \vec{p}_\perp, \partial_z \right) \, f_{p_\perp}\left(\pm z - p_0/eE\right)$ and $f_1 \overleftrightarrow{D_\mu} f_2 = \left(D_\mu f_1\right) f_2 - f_1 \left(D^*_\mu f_2\right)$; $t_0$ is the moment of time after which we adiabatically turn on interactions between charged scalars, $\phi$, and quantum gauge fields, $a_\mu$. In these expressions we neglect the difference between $t_{1,2}$ and $t$ in the limit under consideration. This is mathematically rigorous if $n_{\mu\nu}$ and $\kappa_{\mu\nu}$ have a divergence as $t\to +\infty$ and if we would like to single out only the leading contributions. Otherwise we do such an approximation just to estimate the quantities under consideration.  The physical meaning of such loop corrections is discussed in \cite{Akhmedov:2014hfa}.

Let us consider $n_{\mu\nu}$ in (\ref{nkappa}). In order to estimate the expression in (\ref{nkappa}) we make the change of integration variables to: $t' = \frac{t_3 + t_4}{2}, \tau = t_3 - t_4$. Then, we obtain the $\tau$--integral in the range $[t_0-t, \, t-t_0]$, but its integrand is rapidly oscillating for large $\tau$, as $t\to + \infty$ and $t_0 \to -\infty$. Hence, we can extend the upper and lower limits of the $\tau$--integration to plus and minus infinity, respectively. Then, the integral over $\tau$ leads to the $\delta$--function in the following expression:

\begin{eqnarray}\label{preanswer}
n_{\mu\nu}\left(q_3, q'_3, \vec{q}_\perp, t\right) \approx
e^2 \int_{t_0}^t dt' \int \frac{d^3 \bar{k}}{(2\pi)^3} \int dk'_0 \delta\left(\frac{|q|+|q'|}{2}+k_0'+k_0\right)\frac{e^{- i (|q|-|q'|) t'}}{2 \sqrt{|q||q'|}}\times\notag\\\times \int dz_3 e^{ i q_3 z_3} \left[f_{k_\perp}\left( z_3 - \frac{k_0}{e E}\right)\overleftrightarrow{D_\mu} f_{\left|\vec{q}_\perp + \vec{k}_\perp\right|}\left(-z_3 - \frac{k_0'}{e E}\right)\right]\times\notag\\\times \int dz_4 e^{ - i q'_3 z_4} \left[f^*_{k_\perp}\left(z_4 - \frac{k_0}{e E}\right)\overleftrightarrow{D_\nu} f^*_{\left|\vec{q}_\perp + \vec{k}_\perp\right|}\left(-z_4 - \frac{k_0'}{e E}\right)\right].
\end{eqnarray}
We further make the following change of integration variables $Z = \frac{z_3 + z_4}{2}$ and $z = z_3 - z_4$. Also we change $k_0 \to k_0 - eEZ$ and $k_0' \to k_0' + eEZ $. This change of integration variables allows us to simplify the integral over $Z$, which leads to a $\delta$--function establishing that $q_3 = q_3'$. As a result, $n_{\mu\nu}\left(q_3,q'_3, \vec{q}_\perp, t\right) = \delta\left(q_3 - q'_3\right) \, n_{\mu\nu}\left(\vec{q},t\right)$, where

\begin{eqnarray}\label{answer}
n_{\mu\nu}\left(\vec{q},t\right) \approx e^2 \, (t-t_0) \, \int \frac{d^3 \bar{k}}{(2\pi)^3}\frac{1}{2 |q|}\times \nonumber \\ \times \int dz e^{-i \, 2 \, q_3 \, z} \left[f_{k_\perp}\left(z - \frac{k_0}{e E}\right)\overleftrightarrow{D_\mu} f_{\left|\vec{q}_\perp + \vec{k}_\perp\right|}\left(- z + \frac{k_0+|q|}{e E}\right)\right] \times \nonumber \\ \times \left[f^*_{k_\perp}\left(- z - \frac{k_0}{e E}\right)\overleftrightarrow{D_\nu} f^*_{\left|\vec{q}_\perp + \vec{k}_\perp\right|}\left(z + \frac{k_0+|q|}{e E}\right)\right].
\end{eqnarray}
To obtain this expression from (\ref{preanswer}) we have used that $|q| = |q'|$ due to the presence of $\delta^{(2)}\left(\vec{q}_\perp - \vec{q}'_\perp\right) \, \delta\left(q_3 - q_3'\right)$ in $n_{\mu\nu}\left(\vec{q}, \vec{q}', t\right)$. Also we evaluate the integral over $t'$ in (\ref{preanswer}).

Finally, making the Fourier transformation (\ref{Fourier}), one can straightforwardly see that (\ref{answer}) coincides with the expression for $n_{\mu\nu}$ obtained in \cite{Akhmedov:2014hfa}. Thus, $n_{\mu\nu}$ is divergent as $(t-t_0) \to \infty$. This divergence signals the presence of the photon production which starts right after the moment $t_0$, when the interactions are turned on. It brakes the time reversal and translational invariance of QED on the constant field background. We discuss the physical meaning of all these observations in \cite{Akhmedov:2014hfa} in greater detail, but let us clarify a few points here.

The photon's Keldysh propagator does not depend on $t = t_1 + t_2$ on the tree--level.
But the Keldysh propagator contains such an element as $\langle \alpha^+_\mu \alpha_\nu \rangle$. At the tree--level this quantity does not depend on time, because all the time dependence is absorbed into the harmonics. However, if one turns on selfinteractions of quantum charged fields with quantum photons, this quantity starts to run in the loops. It happens that in the stationary situation it vanishes in the limit $t-t_0 \to \infty$, because of the energy--momentum conservation ($\langle \alpha^+_\mu \alpha^\mu \rangle$ is proportional to the corresponding $\delta$--functions imposing the conservation laws). That should be the case, because in the stationary situation the level--population does not change. However, in the background fields there is no energy conservation or the energy is not bounded from below. As the result $\langle \alpha^+_\mu \alpha^\mu \rangle$ shows the secular growth, i.e. is proportional to $t-t_0$ --- level--populations start to change immediately right after one switches on self--interactions (after $t_0$) and continues till the moment of observation, $t$. The coefficient in front of $(t-t_0)$ is nothing but a part of the collision integral responsible for particle creation by the background field. As the result, unlike the stationary situation, one cannot take the moment $t_0$ (after which the coupling constant is adiabatically turned one) to past infinity.

Let us continue now with the consideration of $\kappa_{\mu\nu}$. In \cite{Akhmedov:2014hfa} we show that it does not receive growing contributions. (This, in particular, shows that the initial state for the photons is the appropriate vacuum state.) Now we are going to show, that in the static gauge, $\kappa_{\mu\nu}$ also does not grow with time.
Similarly to the case of $n_{\mu\nu}$ here we also get that $\kappa_{\mu\nu}(q_3,q_3', \vec{q}_\perp)\propto \delta(q_3 + q_3')$. Then, we take the limit $t\to\infty$ and $t_0 \to -\infty$ in (\ref{nkappa}). This way we find that $\kappa_{\mu\nu}(q_3,\vec{q}_\perp) \propto \delta(k_0 + k_0' + |q|) \delta(k_0 + k_0' - |q|)$. Hence, one can integrate out $k_0'$ to find that:

\begin{gather}
\kappa_{\mu\nu}\left(\vec{q}, t\to + \infty\right) \approx - 2
e^2  \frac{1}{|q|} \, \delta\left(2|q|\right) \, \int \frac{d^3\bar{k}}{(2\pi)^3}
\int dz e^{ i 2 q_3 z} \times \notag \\ \times \left[f_{k_\perp}\left(z - \frac{k_0}{e E}\right)\overleftrightarrow{D_\mu} f_{\left|\vec{q}_\perp + \vec{k}_\perp\right|}\left(-z + \frac{k_0 + |q|}{e E}\right)\right] \, \left[f^*_{k_\perp}\left(z - \frac{k_0}{e E}\right)\overleftrightarrow{D_\nu} f^*_{\left|\vec{q}_\perp + \vec{k}_\perp\right|}\left(-z + \frac{k_0 + |q|}{e E}\right)\right] \nonumber
\end{gather}
The obtained expression contains only convergent integrals and, hence, is finite, if $q\neq 0$.

\subsection{Correction to the Keldysh propagator of the charged particles}

The one--loop correction to the scalar Keldysh propagator, in the limit $t = (t_1 + t_2)/2 \to \infty$ and $t_1 - t_2 = const$, can also be expressed as (\ref{loop2}) where $n^\pm\left(\bar{k}, \bar{k}', t\right) = \delta^{(2)}\left(\vec{k}_\perp - \vec{k}'_\perp\right)\, n^\pm\left(k_0,k'_0, \vec{k}_\perp, t\right)$ and similarly for the case of $\kappa^\pm$. In this case, for example,

\begin{eqnarray}
n^+\left(k_0,k'_0, \vec{k}_\perp, t\right) =
e^2\int \frac{d^3 \vec{q}}{(2\pi)^3} \int \frac{dk_0''}{2\pi}  \int\limits^{t}_{t_0}\int\limits^{t}_{t_0} dt_3 dt_4 \frac{e^{- i \left(|q|+k_0''\right) (t_3-t_4)} e^{- i k_0 t_3 + i k_0' t_4}}{2 |q|}\times\notag\\\times
\int dz_3 e^{ i q_3 z_3} \left[f_{k_\perp}\left(z_3 - \frac{k_0}{e E}\right)\overleftrightarrow{D_\mu} f_{\left|\vec{q}_\perp + \vec{k}_\perp\right|}\left(-z_3 - \frac{k_0''}{e E}\right)\right] \times \notag \\ \times
\int dz_4 e^{ - i q_3 z_4} \left[f^*_{k_\perp}\left(z_4 - \frac{k'_0}{e E}\right)\overleftrightarrow{D_\mu} f^*_{\left|\vec{q}_\perp + \vec{k}_\perp\right|}\left(-z_4 - \frac{k_0''}{e E}\right)\right]\notag\\
{\rm and} \quad \kappa^+ \left(k_0,k'_0,\vec{k}_\perp, t\right) = - 2
e^2\int \frac{d^3 \vec{q}}{(2\pi)^3} \, \int \frac{dk_0''}{2\pi}  \int\limits^{t_1}_{t_0}\int\limits^{t_3}_{t_0 } dt_3 dt_4 \frac{e^{- i \left(|q|+k_0''\right) (t_3-t_4)} e^{- i k_0 t_3 - i k_0' t_4}}{2 q} \times \notag \\ \times
\int dz_3 e^{ i q_3 z_3} \left[f_{k_\perp}\left(z_3 - \frac{k_0}{e E}\right)\overleftrightarrow{D_\mu} f_{\left|\vec{q}_\perp + \vec{k}_\perp\right|}\left(-z_3 - \frac{k_0''}{e E}\right)\right] \times \notag \\ \times
\int dz_4 e^{ - i q_3 z_4} \left[f_{k_\perp}\left(z_4 - \frac{k'_0}{e E}\right)\overleftrightarrow{D_\mu} f^*_{\left|\vec{q}_\perp + \vec{k}_\perp\right|}\left(-z_4 - \frac{k_0''}{e E}\right)\right].
\end{eqnarray}
There are similar expressions for $n^-$ and $\kappa^-$.

In \cite{Akhmedov:2014hfa} we show that none of the $n^\pm$ and $\kappa^\pm$ receive corrections that grow with time. To make the same conclusion here we perform the same trick as at the end of the previous subsection. For example, let us consider $n^+$ and take $t\to + \infty$ and $t_0 \to -\infty$. Then, performing the same transformations as at the end of the previous subsection, we find:

\begin{eqnarray}
n^+\left(k_0,k'_0, \vec{k}_\perp, t\to + \infty\right) \approx
e^2\delta(k_0-k_0')\int \frac{d^3 \vec{q}}{(2\pi)^2}\frac{1}{2 q} \times\notag\\\times
\left|\int dz_3 e^{ i q_3 z_3} \left[f_{k_\perp}\left( z_3 - \frac{k_0}{e E}\right)\overleftrightarrow{D_\mu} f_{\left|\vec{q}_\perp + \vec{k}_\perp\right|}\left(-z_3 + \frac{k_0+|q|}{e E}\right)\right]\right|^2.
\end{eqnarray}
This expression contains only convergent integrals. Hence, $n^+$ cannot contain contributions that grow with time. Using the same line of arguments one can draw the same conclusion for the case of $n^-$ and $\kappa^\pm$.

\section{Discussion}

We would like to present here some additional physical consequences of the observations made above and in our previous paper.

\subsection{Remarks on the loop correction to the current of the created particles}

Since we have shown that the result of \cite{Akhmedov:2014hfa} is gauge independent, we prefer to use the temporal gauge, i.e. $A_\mu = (0,0,0,-Et)$, because then the situation is easier to generalize to more physically natural situations such as the pulse background.

The fact that $n^\pm$ do not grow with time does not necessarily mean that there is no charge particle production generated by loops. First, it is worth stressing here that the correct particle number in the temporal gauge is $n^\pm\left(\vec{k},t\right) \, \left|f_{k_\perp}\left(\pm t + \frac{k_3}{eE}\right)\right|^2$ rather than $n^\pm$ itself. Second, although $n^\pm\left(\vec{k},t\to -\infty \right) = 0$, $\kappa^\pm\left(\vec{k},t\to -\infty\right) = 0$ it is the case that $n^\pm\left(\vec{k},t\to + \infty \right) = n^\pm \neq 0$, $\kappa^\pm\left(\vec{k},t\to + \infty\right) = \kappa^\pm \neq 0$. This kind of behavior of $n^\pm$ and $\kappa^\pm$ is clearly another sign of the breaking of the time translational and reversal invariance of the theory, which is respected at tree--level.

What physical consequences should all this have? In e.g. \cite{Anderson:2013zia}, \cite{Anderson:2013ila} it was shown that the tree--level current of the produced pairs,

\begin{eqnarray}\label{J3}
\left\langle : J_3 :\right\rangle_{tree} = 2 e \, \int \frac{dp_3 \, d^2\vec{p}_\perp}{\left(2\pi\right)^3} \left(p_3 + eEt\right)\left[\left|f_{p_\perp}\left(t + \frac{p_3}{eE}\right)\right|^2 - \frac{1}{2\, \omega_{p_\perp}\left(p_3 + eEt \right)}\right],
\end{eqnarray}
is vanishing. Here $\omega_{p_\perp}\left(p_3 + eEt \right) = \sqrt{m^2 + \vec{p}^2_{\perp} + \left(p_3 + eEt\right)^2}$ and the last term under the integral cancels UV divergent contribution to the current, if it is present (see e.g. \cite{Anderson:2013ila}, \cite{Anderson:2013zia}). To see the vanishing of (\ref{J3}) one has to convert the integration variables $p_3 \to p_{ph} = p_3 + eEt$ and to note that $\left|f_{p_\perp}(p_{ph})\right|^2$ is an even function of $p_{ph}$. Thus, the current vanishes just as a consequence of the time translation and time reversal invariance of the theory in the constant electric field.

At loop order, time translational and reversal invariance is broken. Hence, we can expect that the one--loop correction to the current will be non--vanishing. In fact, the correction is given by

\begin{gather}\label{Jloop}
\left\langle : J_3 : \right\rangle_{loop} = 4 e \int \frac{dp_{ph} \, d^2 \vec{p}_{\perp}}{(2\pi)^3} \left\{n^+_{p_\perp}(p_{ph})  \left|f_{p_\perp}(p_{ph})\right|^2 + {\rm Re}\left[\kappa^+_{p_\perp}(p_{ph})^{\phantom{\frac12}} f^2_{p_\perp}(p_{ph})\right]\right\} \, p_{ph},
\end{gather}
where we denote $n^+\left(\vec{p}, t\right) = n^+_{p_\perp}\left(p_3 + eEt\right) = n^+_{p_\perp}(p_{ph})$ and similarly for $\kappa^+$. Here $n^\pm$ and $\kappa^\pm$ are indeed functions of $p_{ph} = p_3 + eEt$ \cite{Akhmedov:2014hfa}:

\begin{eqnarray}\label{par_den_der}
n^+_{p_\perp}(p_{ph}) \approx \frac{e}{E} \int^{p_{ph}}_{-\infty} dk_{ph} \int \limits^\infty_{-\infty} d\tau \int \frac{d^{3} q}{(2\pi)^{3}} \frac{e^{-2 i |q|\tau}}{2|q|}
\times \nonumber \\ \times \left[f_{p_{\perp}}\left(\tau + \frac{k_{ph}}{eE}\right) \overleftrightarrow{D_\mu} f_{\left|\vec{p}_{\perp} - \vec{q}_{\perp}\right|}\left(\tau + \frac{k_{ph} - q_3}{eE}\right)\right] \,\left[ f^*_{p_{\perp}}\left(\tau - \frac{k_{ph}}{eE}\right) \overleftrightarrow{D_\mu} f^*_{\left|\vec{p}_{\perp}-\vec{q}_\perp\right|}\left(\tau - \frac{k_{ph} - q_3}{eE}\right)\right], \nonumber \\
{\rm and} \quad \kappa^+_{p_\perp}(p_{ph}) \approx - \frac{2\,e}{E} \int^{p_{ph}}_{-\infty} dk_{ph} \int \limits^\infty_{-\infty} d\tau \int \frac{d^{3} q}{(2\pi)^{3}} \frac{e^{-2 i |q|\tau}}{2|q|}
\times \nonumber \\ \times \left[f^*_{p_{\perp}}\left(\tau + \frac{k_{ph}}{eE}\right) \overleftrightarrow{D_\mu} f_{\left|\vec{p}_{\perp} - \vec{q}_{\perp}\right|}\left(\tau + \frac{k_{ph} - q_3}{eE}\right)\right] \,\left[ f^*_{p_{\perp}}\left(\tau - \frac{k_{ph}}{eE}\right) \overleftrightarrow{D_\mu} f^*_{\left|\vec{p}_{\perp}-\vec{q}_\perp\right|}\left(\tau - \frac{k_{ph} - q_3}{eE}\right)\right],
\end{eqnarray}
where $D_\mu \, f_{p_\perp}\left(t + \frac{p_3}{eE}\right) \equiv \left(\partial_t, i p_1, i p_2, i p_3 + i eEt\right)\, f_{p_\perp}\left(t + \frac{p_3}{eE}\right)$. Furthermore, to derive (\ref{Jloop}) we use that $\kappa^-$ is just the complex conjugate of $\kappa^+$ and $n^-\left(p_{ph}\right) = n^+\left(-p_{ph}\right)$, which is straightforward to show.

It is not hard to see that (\ref{Jloop}) is not zero. The point is that $n^+$ and $\kappa^+$ are not even functions of $p_{ph}$. For any choice of the harmonic functions, $f_{k_\perp}$, these quantities do vanish as $p_{ph} \to -\infty$ and approach finite non-zero constants as $p_{ph}\to + \infty$.

In order to estimate (\ref{Jloop}), we note that in--harmonics behave as:

$$f_{p_\perp}(p_{ph}) \propto \left(\frac{p_{ph}}{m}\right)^{i\frac{\vec{p}^2_\perp + m^2}{2 e E}} \frac{\exp\left[i \frac{p_{ph}^2}{2 e E}\right]}{\sqrt{2} \left(m^2 + \vec{p}^2_\perp + p_{ph}^2\right)^{\frac14}},$$
when $p_{ph} \to -\infty$ and

\begin{eqnarray}
f_{p_\perp}(p_{ph}) \approx \alpha_{p_\perp} \cdot \left(\frac{p_{ph}}{m}\right)^{i\frac{\vec{p}^2_\perp + m^2}{2 e E}} \frac{\exp\left[i \frac{p_{ph}^2}{2 e E}\right]}{\sqrt{2} \left(m^2 + \vec{p}^2_\perp + p_{ph}^2\right)^{\frac14}} + \beta_{p_\perp} \cdot \left(\frac{p_{ph}}{m}\right)^{-i\frac{\vec{p}^2_\perp + m^2}{2 e E}}\frac{\exp\left[-i \frac{p^2_{ph}}{2 e E}\right]}{\sqrt{2}\left(m^2 + \vec{p}^2_\perp + p_{ph}^2\right)^{\frac14}},\nonumber
\end{eqnarray}
when $p_{ph} \to +\infty$. Here  $\alpha_{p_\perp}$ and $\beta_{p_\perp}$ are functions of $p_\perp$, obeying the condition $|\alpha_{p_\perp}|^2 - |\beta_{p_\perp}|^2 = 1$. Then, defining $n^+_{p_\perp}\left(p_{ph} = + \infty \right) = n^+_{p_\perp}$, $\kappa^+_{p_\perp}\left(p_{ph} = + \infty\right) = \kappa^+_{p_\perp}$ and using the same approximations as in \cite{PolyakovKrotov}, we obtain

\begin{gather}
\left\langle : J_3 : \right\rangle_{loop} \propto e E^2 (t-t_0) \int d^2\vec{p}_\perp\left\{n^+_{p_\perp} \left|\beta_{p_\perp}\right|^2 + {\rm Re}\left[\kappa^+_{p_\perp} \, \alpha_{p_\perp} \, \beta_{p_\perp}\right]\right\},
\end{gather}
This expression is similar to the one obtained in \cite{Gavrilov:2007hq}, \cite{Akhmedov:2009vs}, \cite{PolyakovKrotov}, \cite{Anderson:2013ila}, \cite{Anderson:2013zia} in the pulse background. The crucial difference with the tree--level result for the pulse background, however, comes from the fact that $n^+ \sim e^2$ and $\kappa^+ \sim e^2$ are the results of the one--loop contribution.

\subsection{Remarks on the solution of the kinetic equation and summation of the leading loop corrections}

In \cite{Akhmedov:2014hfa} we show that $n_{\mu\nu}$ for the photons is equal to $n_{\mu\nu}\left(\vec{q}, t\right) = \pi_{\mu\nu} n_q\left(t\right)$, where $\pi_{\mu\nu}$ is time independent, symmetric, transversal, $q^\mu \pi_{\mu\nu} = 0$, $q^2 = 0$, tensor. Then from the system of Dyson-Schwinger equations we derive a kinetic equation for $n_q\left(t\right)$:

\begin{equation}\label{simpl_kin_eq}
\frac{\partial n_q(t)}{\partial t} = \Gamma_1(q) \left[1 + n_q(t)\right] - \Gamma_2(q)\, n_q(t),
\end{equation}
where

\begin{eqnarray}\label{kin_eq}
\Gamma_1(q) \approx e^2 \int \frac{d^3 k}{(2\pi)^3} \int\limits^\infty_{-\infty} d\tau \frac{e^{-2 i |q| \tau}}{|q|} \,\left[f_{k_\perp}\left(\tau + \frac{k_3}{eE}\right) \overleftrightarrow{D_\mu} f_{\left|\vec{k}_\perp - \vec{q}_\perp\right|}\left(\tau + \frac{k_3 - q_3}{eE}\right)\right] \times \nonumber \\ \times \left[ f^*_{k_\perp}\left(\tau - \frac{k_3}{eE}\right) \overleftrightarrow{D_\mu} f^*_{\left|\vec{k}_\perp - \vec{q}_\perp\right|}\left(\tau - \frac{k_3 - q_3}{eE}\right)\right] \notag \\
{\rm and} \quad \Gamma_2(q) \approx e^2 \int \frac{d^3 k}{(2\pi)^3} \int\limits^\infty_{-\infty} d\tau \frac{e^{-2 i |q| \tau}}{|q|} \, \left[f^*_{k_\perp}\left(\tau + \frac{k_3}{eE}\right) \overleftrightarrow{D_\mu} f^*_{\left|\vec{k}_\perp - \vec{q}_\perp\right|}\left(\tau + \frac{k_3 - q_3}{eE}\right)\right] \times \nonumber \\ \times \left[f_{k_\perp}\left(\tau - \frac{k_3}{eE}\right) \overleftrightarrow{D_\mu} f_{\left|\vec{k}_\perp - \vec{q}_\perp\right|}\left(\tau - \frac{k_3 - q_3}{eE}\right)\right].
\end{eqnarray}
The physical meaning of (\ref{simpl_kin_eq}) is transparent. The first term on the right hand side describes the photon production by the background field, while the second term accounts for the decay of the produced photons into charged pairs. These processes are allowed in the presence of the background field. The absence of other terms describing other processes is explained by their suppression by higher powers of $e^2$ \cite{Akhmedov:2014hfa}. The solution of (\ref{simpl_kin_eq}) sums up leading corrections, i.e. unsuppressed powers of $e^2(t-t_0)$, from all loops. Here we would like to find/compare $\Gamma_1$ and $\Gamma_2$ and, hence, to solve this kinetic equation.

To find the relation between $\Gamma_1$ and $\Gamma_2$, note that
generic harmonic functions look like (see e.g. \cite{Anderson:2013zia}, \cite{Anderson:2013ila}):

\begin{equation}
f_{k_\perp}\left(t + \frac{k_3}{e E}\right) = A \, D_{-\frac{1}{2}+ i \frac{m^2 + k^2_\perp}{2 e E}} \left[- e^{-i\frac{\pi}{4}} \sqrt{\frac{2}{e E}}\left(k_3 + eEt\right)\right] + B \, D_{-\frac{1}{2} - i \frac{m^2 + k^2_\perp}{2 e E}}\left[- e^{i\frac{\pi}{4}} \sqrt{\frac{2}{e E}}\left(k_3 + eEt\right)\right].
\end{equation}
Where $A$ and $B$ some constants. For example, for the in--harmonics $B=0$. Then, one can see that $f^*_{k_\perp}\left(t + \frac{k_3}{e E}\right)$ is equal to $f_{k_\perp}\left(t + \frac{k_3}{e E}\right)$ under the exchange of $eE \to - eE$ and $\vec{k} \to - \vec{k}$. Using this relation and the change of $\vec{k} \to \vec{q} - \vec{k}$ under the integrals in (\ref{kin_eq}), one can show that $\Gamma_1 = \Gamma_2$. The same is also true for the case of out--harmonics. As a result, for such a choice of the harmonic functions, the leading one--loop correction to $n_{\mu\nu}(\vec{q}, t)$ is exact and we have the linear growth in all loops. This means that the time translational and reversal invariance cannot be restored after summation of all loops.

\section{Acknowledgements}

Our work was partially supported by the grant for the support of the leading scientific schools SSch--1500.2014.2 and by our grants from the Dynasty foundation. The work of FKP is done under the partial support of the RFBR grant 14-02-31446-mol-a. The work of ETA was done under the financial support from the Government of the Russian Federation within the framework of the implementation of the 5-100 Programme Roadmap of the National Research University Higher School of Economics.


\thebibliography{50}

\bibitem{Schw}
  J.~S.~Schwinger,
  Phys.\ Rev.\  {\bf 82}, 664 (1951).

\bibitem{Akhmedov:2014hfa}
  E.~T.~Akhmedov, N.~Astrakhantsev and F.~K.~Popov,
  JHEP {\bf 1409}, 071 (2014)
  [arXiv:1405.5285 [hep-th]].

\bibitem{Akhmedov:2009vh}
  E.~T.~Akhmedov and E.~T.~Musaev,
  ``Comments on QED with background electric fields,''
  New J.\ Phys.\  {\bf 11}, 103048 (2009)
  [arXiv:0901.0424 [hep-ph]].

\bibitem{Gitman1}
  E.~S.~Fradkin and D.~M.~Gitman,
  Fortsch.\ Phys.\  {\bf 29}, 381 (1981).

\bibitem{Gitman:1986xr}
  D.~M.~Gitman, E.~S.~Fradkin and S.~M.~Shvartsman,
  Fortsch.\ Phys.\  {\bf 36}, 643 (1988).

\bibitem{Gavrilov:1980cs}
  S.~P.~Gavrilov, D.~M.~Gitman and S.~M.~Shvartsman,
  Sov.\ Phys.\ J.\  {\bf 23}, 257 (1980).

\bibitem{Nikishov}
  N.~B.~Narozhnyi and A.~I.~Nikishov,
  Teor.\ Mat.\ Fiz.\  {\bf 26}, 16 (1976).

\bibitem{Nikishov:1974zv}
  A.~I.~Nikishov,
  Teor.\ Mat.\ Fiz.\  {\bf 20}, 48 (1974).

\bibitem{Nikishov:1969tt}
  A.~I.~Nikishov,
  Zh.\ Eksp.\ Teor.\ Fiz.\  {\bf 57}, 1210 (1969).

\bibitem{Gitman2}
  D.~M.~Gitman and S.~P.~Gavrilov,
  Izv.\ Vuz.\ Fiz.\  {\bf 1}, 94 (1977)

\bibitem{Gavrilov:1979bj}
  S.~P.~Gavrilov, D.~M.~Gitman and S.~M.~Shvartsman,
  Yad.\ Fiz.\  {\bf 29}, 1097 (1979).

\bibitem{Volfengaut:1981ea}
  Yu.~Y.~Volfengaut, S.~P.~Gavrilov, D.~M.~Gitman and S.~M.~Shvartsman,
  Yad.\ Fiz.\  {\bf 33}, 743 (1981).

\bibitem{Gavrilov:1982we}
  S.~P.~Gavrilov and D.~M.~Gitman,
  Sov.\ Phys.\ J.\  {\bf 25}, 775 (1982).

\bibitem{Gavrilov:1996pz}
  S.~P.~Gavrilov and D.~M.~Gitman,
  Phys.\ Rev.\  D {\bf 53}, 7162 (1996)
  [arXiv:hep-th/9603152].

\bibitem{Gavrilov:2007hq}
  S.~P.~Gavrilov and D.~M.~Gitman,
  Phys.\ Rev.\  D {\bf 78}, 045017 (2008)
  [arXiv:0709.1828 [hep-th]].

\bibitem{Tomaras:2000ag}
  T.~N.~Tomaras, N.~C.~Tsamis and R.~P.~Woodard,
  Phys.\ Rev.\ D {\bf 62}, 125005 (2000)
  [hep-ph/0007166].

\bibitem{Cooper:1989kf}
  F.~Cooper and E.~Mottola,
  Phys.\ Rev.\ D {\bf 40}, 456 (1989).

\bibitem{Cooper:1987pt}
  F.~Cooper and E.~Mottola,
  Phys.\ Rev.\ D {\bf 36}, 3114 (1987).

\bibitem{Kluger:1998bm}
  Y.~Kluger, E.~Mottola and J.~M.~Eisenberg,
  Phys.\ Rev.\ D {\bf 58}, 125015 (1998)
  [hep-ph/9803372].

\bibitem{Cooper:1992hw}
  F.~Cooper, J.~M.~Eisenberg, Y.~Kluger, E.~Mottola and B.~Svetitsky,
  Phys.\ Rev.\ D {\bf 48}, 190 (1993)
  [hep-ph/9212206].

\bibitem{Kluger:1992gb}
  Y.~Kluger, J.~M.~Eisenberg, B.~Svetitsky, F.~Cooper and E.~Mottola,
  Phys.\ Rev.\ D {\bf 45}, 4659 (1992).

\bibitem{Kluger:1991ib}
  Y.~Kluger, J.~M.~Eisenberg, B.~Svetitsky, F.~Cooper and E.~Mottola,
  Phys.\ Rev.\ Lett.\  {\bf 67}, 2427 (1991).

\bibitem{Gelis:2013oca}
  F.~Gelis and N.~Tanji,
  Phys.\ Rev.\ D {\bf 87}, no. 12, 125035 (2013)
  [arXiv:1303.4633 [hep-ph]].

\bibitem{Fukushima:2009er}
  K.~Fukushima, F.~Gelis and T.~Lappi,
  Nucl.\ Phys.\ A {\bf 831}, 184 (2009)
  [arXiv:0907.4793 [hep-ph]].

\bibitem{Karbstein:2013ufa}
  F.~Karbstein,
  Phys.\ Rev.\ D {\bf 88}, no. 8, 085033 (2013)
  [arXiv:1308.6184 [hep-th]].

\bibitem{Dunne:2005sx}
  G.~V.~Dunne and C.~Schubert,
  Phys.\ Rev.\ D {\bf 72}, 105004 (2005)
  [hep-th/0507174].

\bibitem{Dunne:2006ff}
  G.~V.~Dunne and C.~Schubert,
  AIP Conf.\ Proc.\  {\bf 857}, 240 (2006)
  [hep-ph/0604089].

\bibitem{Dunne:2006st}
  G.~V.~Dunne, Q.~-h.~Wang, H.~Gies and C.~Schubert,
  Phys.\ Rev.\ D {\bf 73}, 065028 (2006)
  [hep-th/0602176].

\bibitem{Schubert:2007xm}
  C.~Schubert,
  AIP Conf.\ Proc.\  {\bf 917}, 178 (2007)
  [hep-th/0703186].

\bibitem{Ruffini:2003cr}
  R.~Ruffini, L.~Vitagliano and S.~S.~Xue,
  Phys.\ Lett.\ B {\bf 559}, 12 (2003)
  [astro-ph/0302549].

\bibitem{Grib} Grib A. A., Mamaev S. G., Mostepanenko V. M. ``Quantum effects in strong external
fields'', Atomizdat, Moscow 1980, 296.\\ Grib A. A., Mamayev S. G., Mostepanenko V. M. Vacuum quantum effects in strong fields. – St. Petersburg : Friedmann Laboratory, 1994.

\bibitem{LL}
L.~D.~Landau and E.~M.~Lifshitz, Vol. 10 (Pergamon Press, Oxford, 1975).

\bibitem{Kamenev} A.Kamenev, ``Many-body theory of non-equilibrium systems'',  arXiv:cond-mat/0412296;
Bibliographic Code:	2004cond.mat.12296K.

\bibitem{Barashev:1985zm}
  V.~P.~Barashev, A.~E.~Shabad and S.~M.~Shvartsman,
  Sov.\ J.\ Nucl.\ Phys.\  {\bf 43}, 617 (1986)
  [Yad.\ Fiz.\  {\bf 43}, 964 (1986)].

\bibitem{Anderson:2013zia}
  P.~R.~Anderson and E.~Mottola,
  arXiv:1310.1963 [gr-qc].

\bibitem{Anderson:2013ila}
  P.~R.~Anderson and E.~Mottola,
  arXiv:1310.0030 [gr-qc].

\bibitem{PolyakovKrotov}
  D.~Krotov, A.~M.~Polyakov,
  Nucl.\ Phys.\  {\bf B849}, 410-432 (2011).
  [arXiv:1012.2107 [hep-th]].


\bibitem{Akhmedov:2008pu}
 E.~T.~Akhmedov and P.~V.~Buividovich,
  ``Interacting Field Theories in de Sitter Space are Non-Unitary,''
 Phys.\ Rev.\  D {\bf 78}, 104005 (2008)
  [arXiv:0808.4106 [hep-th]].

\bibitem{Akhmedov:2009be}
  E.~T.~Akhmedov, P.~V.~Buividovich and D.~A.~Singleton,
  Phys.\ Atom.\ Nucl.\  {\bf 75}, 525 (2012)
  [arXiv:0905.2742 [gr-qc]].

\bibitem{AkhmedovKEQ}
  E.~T.~Akhmedov,
  JHEP {\bf 1201}, 066 (2012)
  [arXiv:1110.2257 [hep-th]].

\bibitem{AkhmedovBurda}
  E.~T.~Akhmedov and P.~Burda,
  Phys.\ Rev.\ D {\bf 86}, 044031 (2012)
  [arXiv:1202.1202 [hep-th]].

\bibitem{Polyakov:2012uc}
  A.~M.~Polyakov,
  ``Infrared instability of the de Sitter space,''
  arXiv:1209.4135 [hep-th].

\bibitem{AkhmedovGlobal}
  E.~T.~Akhmedov,
  Phys.\ Rev.\ D {\bf 87}, 044049 (2013)
  [arXiv:1209.4448 [hep-th]].

\bibitem{AkhmedovPopovSlepukhin}
  E.~T.~Akhmedov, F.~K.~Popov and V.~M.~Slepukhin,
  Phys.\ Rev.\ D {\bf 88}, 024021 (2013)
  [arXiv:1303.1068 [hep-th]].

\bibitem{AkhmedovReview}
  E.~T.~Akhmedov,
  Int.\ J.\ Mod.\ Phys.\ D {\bf 23}, no. 1, 1430001 (2014)
  [arXiv:1309.2557 [hep-th]].

\bibitem{Akhmedov:2009vs}
  E.~T.~Akhmedov and P.~Burda,
  ``A Simple way to take into account back reaction on pair creation,''
  Phys.\ Lett.\ B {\bf 687}, 267 (2010)
  [arXiv:0912.3435 [hep-th]].
 \end{document}